\input harvmac
\input epsf
\noblackbox
\def\xt{\tilde{x}}

\def\zt{\tilde{z}}

\def\zh{\hat{z}}

\def\dwn{\Delta^*_{W_{n+1}}}

\def\nus{\nu^\star}

\def\hx#1{{\hat{#1}}}

\def\Ds{\Delta^\star}
\def\abstract#1{
\vskip .5in\vfil\centerline
{\bf Abstract}\penalty1000
{{\smallskip\ifx\answ\bigans\leftskip 2pc \rightskip 2pc
\else\leftskip 5pc \rightskip 5pc\fi
\noindent\abstractfont \baselineskip=12pt
{#1} \smallskip}}
\penalty-1000}
\def\us#1{\underline{#1}}
\def\hth/#1#2#3#4#5#6#7{{\tt hep-th/#1#2#3#4#5#6#7}}
\def\nup#1({Nucl.\ Phys.\ $\us {B#1}$\ (}
\def\plt#1({Phys.\ Lett.\ $\us  {B#1}$\ (}
\def\cmp#1({Comm.\ Math.\ Phys.\ $\us  {#1}$\ (}
\def\prp#1({Phys.\ Rep.\ $\us  {#1}$\ (}
\def\prl#1({Phys.\ Rev.\ Lett.\ $\us  {#1}$\ (}
\def\prv#1({Phys.\ Rev.\ $\us  {#1}$\ (}
\def\mpl#1({Mod.\ Phys.\ Let.\ $\us  {A#1}$\ (}
\def\ijmp#1({Int.\ J.\ Mod.\ Phys.\ $\us{A#1}$\ (}
\def\br{\hfill\break}\def\ni{\noindent}
\def\cx#1{{\cal #1}}\def\IP{{\bf P}}
\def\tx#1{{\tilde{#1}}}

\lref\FMW{R. Friedman, J.W. Morgan and E. Witten, \cmp 187 (1997) 679.}
\lref\MV{C. Vafa and D. Morrison, \nup 473 (1996) 74; \nup 476 (1996) 437.}
\lref\BM{P. Berglund and P. Mayr, {\it Heterotic string/F-theory duality
from mirror symmetry}, hep-th/9811217.}
\lref\FF{P. Mayr, \nup 494 (1997) 489.}
\lref\VW{C. Vafa and E. Witten, Nucl. Phys. Proc. Suppl. {$\us {46}$} 
(1996) 225.} 
\lref\KMV{S. Katz, P. Mayr and C. Vafa,
Adv. Theor. Math. Phys. $\us {1}$ (1998) 53.}
\lref\SDS{A. Klemm, 
W. Lerche, P. Mayr, C. Vafa, N. Warner,
                \nup 477 (1996) 746.}
\lref\KKV{S. Katz, A. Klemm and C. Vafa, \nup 497 (1997) 173.}
\lref\Loo{E. Looijenga, Invent. Math. {$\us {38}$} (1977) 17;
Invent. Math. {$\us {61}$} (1980) 1.}
\lref\zerotwo{E. Witten, \nup 403 (1993) 159; \br
J. Distler and S. Kachru, \nup 413 (1994) 213.}

\Title{\vbox{
\rightline{\vbox{\baselineskip12pt
\hbox{CERN-TH/99-102}
\hbox{NSF-ITP-99-23}
\hbox{hep-th/9904114}}}}}
{Stability of Vector Bundles}
\vskip-1cm\centerline{{\titlefont From F-theory}}\vskip 0.3cm
\centerline{P. Berglund\foot{berglund@itp.ucsb.edu} and  
P. Mayr\foot{Peter.Mayr@cern.ch}}
\vskip 0.6cm
\centerline{$^1$ \it Institute for Theoretical Physics, 
University of California, Santa Barbara, CA 93106, USA}
\vskip 0.0cm
\centerline{$^2$ \it Theory Division, CERN, 1211 Geneva 23, 
Switzerland}
\vskip -0.8cm
\abstract{\ni
{}We use a recently proposed formulation of stable holomorphic vector
bundles $V$ on elliptically fibered Calabi--Yau $n$-fold $Z_n$ in terms 
of toric geometry to describe stability conditions on $V$. Using 
the toric map $f: \ W_{n+1} \to (V,Z_n)$ that identifies dual pairs
of F-theory/heterotic duality we show how stability can be 
related to the
existence of holomorphic sections of a certain line bundle that is
part of the toric construction.}
\Date{\vbox{\hbox{\sl {April 1999}}
}}
\goodbreak

\parskip=4pt plus 15pt minus 1pt
\baselineskip=15pt plus 2pt minus 1pt

The heterotic string compactified on Calabi--Yau $n$ fold $Z_n$
gives rise to interesting minimal supersymmetric string vacua with a
quite realistic spectrum of gauge symmetries and matter fields
\ref\CHSW{P. Candelas, G.T. Horowitz, A. Strominger and
E. Witten, \nup 258 (1985) 46;\br
C. Hull and E. Witten, \plt 160 (1985) 398.;\br
E. Witten, \nup 268 (1986) 79;\br
J. Distler,  \plt 188 (1987) 431;\br
J. Distler and B. Greene, \nup 304 (1988) 1.}.
The definition of these vacua, involves the 
specification of a gauge background $V$ with structure group $H\in G_0$ 
embedded in the perturbative heterotic gauge group $G_0=E_8\times E_8$ or
$SO(32)$ such that
\eqn\vtop{
c_1(V) = 0\ {\rm mod}\ 2,\qquad \lambda(V) = c_2(Z)+[W],}
where $\lambda$ is the four-dimensional characteristic class of
$V$. The second term $[W]$ applies to a generalization of heterotic vacua
which include non-perturbative five-branes 
\ref\Ruben{M.J. Duff, R. Minasian and E. Witten,
\nup 465 (1996) 413.}. Here $[W]$ is a sum of formal
four-forms that integrate to one in the transverse direction to the
five-branes. Note that the five-branes necessarily have to wrap two-cycles
in a 3-fold compactification; such cycles have to be holomorphic in order
to preserve supersymmetry 
\ref\SSC{K. Becker, M. Becker and 
A. Strominger, \nup 456 (1995) 130; \br
M. Bershadsky, V. Sadov and C. Vafa, \nup 463 (1996) 420.}. 

In addition, the connection on $V$ has to satisfy
\eqn\fsv{
F_{ab}=F_{\bar{a}\bar{b}}=0,\qquad g_{a\bar{b}}F^{a\bar{b}} = 0.}
The first equation says that $V$ is holomorphic, while the second 
equation is interpreted as some stability condition on $V$. 
Although an existence theorem exists under certain conditions
\ref\UY{K. Uhlenbeck and S.T. Yau, {\it On the existence of 
hermitian Yang--Mills connections in stable vector bundles},
preprint (1986).},
explicit solutions to \fsv\ have been known only in very special 
constructions. The difficulty to find appropriate backgrounds $V$
has hampered the study of interesting four-dimensional vacua for a long time.

An entirely independent construction of holomorphic, 
stable vector bundles\foot{We will loosely call $V$ a bundle in the following, 
though our construction 
naturally yields sheaf generalizations of bundles, which provide the relevant
set up for heterotic string compactifications 
\ref\DGM{J. Distler, B.R. Greene and D.R. Morrison, \nup 481 (1996) 289.}.}
$V$
on Calabi--Yau $n$-folds $Z_n$ has been developed in \BM\ using a 
certain limit of type IIA strings compactified on a Calabi--Yau 
$n+1$ fold $W_{n+1}$. In this paper we will obtain a simple
stability criterion for $V$ from the geometrical type II setup.
The relation of type IIA strings and holomorphic
stable vector bundles can be traced back to a 
simple extension of mirror symmetry
in a $K3\times T^2$ compactification \KMV, more specifically a 
discrete symmetry in the moduli space of this compactification 
that relates geometric deformations of K3 to Wilson lines on the $T^2$ 
factor. In other words, the physics of 
Wilson lines on $Z_1=T^2$, is the same as the physics of a type IIA string
compactified on a certain limit $\cx W_2$ of an elliptically K3 
manifold $W_2$. The relevant type IIA
compactification geometry $\cx W_2$ in this limit 
is the local mirror of a K\"ahler
resolution of an $H$ singularity of an elliptically fibered ALE space.
Here $H$ is the structure group of the bundle $V$, as before.
Fibering this picture, using the adiabatic argument of \VW, 
one obtains an equivalence between a type IIA
compactification on (the geometric limit $\cx W_{n+1}$ of) an 
elliptic and K3 fibered Calabi--Yau manifold 
$W_{n+1}$ and holomorphic stable vector bundles on an elliptically
fibered manifold $Z_n$. Holomorphic stable bundles $V$
on smooth elliptic fibrations $Z_n$ have been first considered in \FMW\
and subsequently in 
\ref\BJPS{M. Bershadsky, A. Johansen, T. Pantev and V. Sadov,
\nup 505 (1997) 165;\br
R.Y. Donagi, Asian J. Math. $\us 1$ (1997) 214;
MSRI pub. 28 (1992) 65;\br
G. Curio and R.Y. Donagi, \nup 518 (1998) 603.}.

The construction in \BM\ has been
formulated in terms of toric geometry which for many purposes
is the most useful and most general definition for the Calabi--Yau
manifolds $\cx W_{n+1}$ and $Z_n$. 
The toric construction of the pair $(V,Z_n)$ in terms of an $n+1$
dimensional non-compact Calabi--Yau manifold $\cx W_{n+1}$ works for
any structure group $H$ of the bundle. If $H$ fits
into the perturbative heterotic gauge group $G_0=E_8\times E_8$ or $SO(32)$,
we can interpret $(V,Z_n)$ as a valid perturbative heterotic vacuum.
Precisely in this case one can find an embedding of the geometry $\cx W_{n+1}$ 
into a {\it compact} Calabi--Yau manifold $W_{n+1}$. The statement that
type IIA string compactified on a limit of $W_{n+1}$ is equivalent to
the heterotic string compactification on $(V,Z_n)$ then establishes 
F-theory/heterotic duality \ref\vafaf{C. Vafa, \nup 469 (1996) 403}\MV\ 
in the classical limit. 

Toric geometry describes a Calabi--Yau
$n$-fold $Z_n$ in terms of a convex, toric polyhedron $\dwn$, which is
the convex hull of a set of integral vertices $\nus_i$ in a standard
integral lattice $N$\foot{See \BM\ for a list of reviews on toric geometry in 
the physics literature.}. It is satisfying to observe how the heterotic
physics such as the compactification manifold $Z_n$ and some defining 
data  of the bundle $V$ have a very simple representation in terms of
the polyhedron $\dwn$. Moreover also non-perturbative dynamics of
the heterotic vacuum for singular configurations, in particular 
non-perturbative gauge symmetries and non-perturbative five-branes that 
appear in \vtop,
can be read of from the toric data of $W_{n+1}$ in a simple way.
It is the purpose of this note to extend the dictionary  
between physical and
toric data developed in \BM\ to 
describe stability of the bundle $V$ in terms of convexity
of the polyhedron $\dwn$.

In particular it was noted in \BM, that for a given
structure group $H$ of the bundle, convexity
of the toric polyhedron $\Ds_{\cx W_3}$ 
translates to a bound on the first Chern class
$\eta=c_1(\cx N)$ of a line bundle $\cx N$ that is an important 
characteristic of the bundle $V$. Observing that in six dimensions, 
the bound on $\eta$ from convexity of $\Ds$ translates to the stability
of instantons on K3 with structure group $H$, these bounds were interpreted
as a similar stability condition in lower dimensions.
A general formula for the stability of $V$ in terms of bounds on $\eta$  
was subsequently 
conjectured in ref. \ref\Raj{G. Rajesh,
{\it Toric geometry and F theory / heterotic duality in four-dimensions},
hep-th/9811240.}
which was however not in agreement\foot{The
discriminant criterion used in \Raj\ is not restrictive enough to fix the structure
group $H$.} with the original bounds in \BM.
In the following we derive a general stability bound for $\eta$ from
the toric representation of data $(V,Z_n)$ which is
consistent and, we believe, gives a correct treatment.

We start with a sketch of the toric description of the pair $(V,Z_n)$ in terms 
of the non-compact Calabi--Yau $n+1$-fold $\cx W_{n+1}$. The latter is
elliptically fibered with base $\tilde{B}_n$ and ALE fibered with 
base $B_{n-1}$. The manifold $\cx W_{n+1}$  
is defined as the vanishing locus of a polynomial 
$p$ in a toric ambient space with coordinates $(y,x,\zt,v,x_i)$, 
where $x_i$ denote the coordinates of the base $B_{n-1}$ and $(y,x,\zt,v)$
are certain coordinates on the ALE fiber. The general form of $p$ is 
\eqn\pdef{
p=p_0+p_+=p_0(y,x,\zt;x_i)+\sum_{j=1}^{J} v^jp_+^j(y,x,\zt;x_i)\ ,}
where $p_0$ and $p_+^j$ are quasi-homogeneous polynomials. 
In particular the $v$ independent piece $p_0$ describes
an elliptically fibered Calabi--Yau manifold $Z_n:\ p_0=0$, while 
the polynomials $p^j_+$ contain the information about the bundle $V$. 
More specifically, each node in the affine Dynkin diagram $\Gamma(H)$ with
Dynkin index $s_i$ contributes one monomial to the polynomial $p_+^{s_i}$;
in particular $J=max(s_i)$.
E.g. in the case $H=SU(N)$, we have $N$ nodes of index 1, each of
which contributes a monomial to $p_+^1$. 
The precise form of the 
$v$ dependent part of $p$ is $p_+=v(a_N(x_i)\zt^N+a_{N-2}(x_i)\zt^{N-2}x+\dots+
a_0(x_i)x^{N/2})$ for
$N$ even with the last term being $\sim yx^{(N-3)/2}$ for $N$ odd. The 
complex parameters multiplying the monomials in 
$p_+$ give a projective parametrization of the
moduli space of the $SU(N)$ bundle.

Thus the structure group $H$ determines 
the $y,x,\zt,v$ content of the monomials
appearing in the defining equation $p$ of $\cx W_{n+1}$. It does not 
determine the dependence on the base variables $x_i$, however. This
dependence specifies part the topological class of $V$ that enter
the higher Chern classes $c_2(V),c_3(V),\dots$. However note
that this dependence is quite restricted because of the quasi-homogeneousness
of $p$. It was shown in \BM\ that the dependence of the monomials
in $p$ on $x_i$ is completely determined by specifying the action of two
$C^*$ actions acting on $(y,x,\zt,v,x_i)$. In other words, we have to specify
two line bundles $\cx L$ and $\cx M$ on the base $B_{n-1}$ and the 
coordinates  $(y,x,\zt,v,x_i)$ transform as certain sections of 
$\cx M$ and $\cx L$ such that 
\eqn\sects{
y\sim \cx M^3 \cx L^3,\quad  x\sim \cx M^2 \cx L^2,\quad  \zt\sim \cx M,\quad 
v \sim \cx M^{5-N},\quad  f_{c,d} \sim \cx M^d \cx L^c,}
where $f_{c,d}=f_{c,d}(x_i)$ is the base dependent part in a
monomial $$y^{(6-2b-c)/3}x^b\zt^cv^df_{c,d}(x_i)$$  of $p$. Moreover $N$ is
the highest power of $\zt$ that appears in $p_+^1$. 

{}From the fact that $p_0=0$ describes a Calabi--Yau manifold it follows
that the bundle $\cx L$ is actually the anti-canonical bundle of $B_{n-1}$.
On the other hand $\cx M$ is an intrinsic property of $V$ alone. 
In the following we will derive a bound on the first Chern class 
$c_1(\cx N)$ of the line bundle $\cx N=\cx M \cx L^6$ 
from convexity of the polyhedron $\dwn$. The change from $\cx M$ to 
$\cx N$ is intended to make contact with the notation of ref. \FMW,
where $V$ has been described in terms of sections of a certain weighted
projective bundle $\cx W$ on $B_{n-1}$ with coordinates on $\cx W$ being
sections of $\cx N^{s_i} \cx L^{-d_j}$. Here $s_i$ are the Dynkin indices
as above and $d_i$ are the degrees of the independent Casimir invariants of
$H$. The dependence of higher Chern classes of $V$ on $c_1(\cx N)$ 
has been determined in \FMW\ref\BA{B. 
Andreas, {\it On vector bundles and chiral matter 
in N=1 heterotic compactifications}, hep-th/9802202;
G. Curio, \plt 435 (1998) 39.}.

To keep the discussion as basic as possible we will work directly 
with the defining polynomial $p$ rather than with the toric polyhedron 
$\Ds_{\cx W_{n+1}}$
that determines $p$ 
\ref\batms{V. Batyrev, Duke Math. Journ. $\us {69}$ (1993) 349.}.
In particular, if we consider a structure 
group $H\in G_0$, we have a compact embedding $\cx W_{n+1}\to W_{n+1}$.
Moreover the K3 fiber (that, in a local patch, contains 
the ALE fiber of $\cx W_{n+1}$) has
a singularity $G$ which is the commutant of $H$ in $G_0$ \MV. The defining
polynomial $\hx p$ of $W_{n+1}$ can be written in 
generalized Weierstrass form:
\eqn\gwsf{
\hx p=y^2+x^3+yx\zh a_1+x^2\zh ^2a_2+y\zh ^3a_3+x\zh ^4a_4+\zh ^6a_6\ . }
Note that we
use $(y,x,\zh)$ and $(y,x,\zt)$ to denote the homogeneous coordinates 
of the elliptic fiber of the $n+1$-dimensional Calabi--Yau $W_{n+1}$ 
and the $n$ dimensional Calabi--Yau $Z_n$, respectively.
The $a_n$ are functions of the coordinates $\tilde{x}_i$ of the 
base $\tilde{B}_n$ of the {\it elliptic} fibration of $W_{n+1}$. 
In particular having a singularity of type $G$ above a locus, say at
$z\equiv \xt_1=0$ on the base $\tilde{B}_n$ 
can  be phrased in terms of the behavior
of the $a_n$ near $z=0$, $a_n \sim z^{\delta_n}$ using
Tate's algorithm \ref\Betal{M. Bershadsky et al., \nup 481 (1996) 215.}.

To reiterate, the structure group $H$ determines the singularity $G$
of the K3 fiber of $W_{n+1}$. In turn $G$ is determined by the behavior
of the coefficient functions $a_n$ in the generalized Weierstrass form
$\hx p$ near the singularity $z=0$. Moreover we are interested in
the singularity of the K3 fiber $W_2$ of the K3 fibration 
$W_2\to W_{n+1}\to B_{n-1}$. Thus 
$z$ is now a coordinate on
the base $\IP^1$ of the elliptically fiberered K3, $\pi:W_2\to \IP^1$.
However note that in the local limit
$W_{n+1} \to \cx W_{n+1}$ where our map $f:\ \cx W_{n+1} \to (V,Z_n)$
applies, the $a_n$ become nothing but the $\tx B_n$ dependent
parts of the monomials  $y^{(6-b-c)/3}x^b\zt^cv^df_{c,d}(x_i)$ in
the defining polynomial $p$ of $\cx W_{n+1}$. 

To be more explicit let
us consider the case $G_0=E_8\times E_8$. The relation between the 
coordinates in $\hx p$ and $p$, using $z\equiv \xt_1,\  w \equiv \xt_2$ as
the coordinates of the base $\IP^1$ of $W_2$ as above,
is then 
$$
(y,x,\zt;v;x_i)_p = (y,x,\zh zw;w/z;\xt_i,\ i>3)_{\hx p}.
$$

In particular the 
existence of a term $a_n=z^{\delta_n}f(x_i)$ implies that the line
bundle 
\eqn\lbi{
\cx N^{d_n} \cx L^{n-6d_n}}
has a holomophic section, where we have defined $d_n=n-\delta_n$. 
Combining the conditions imposed by all the $a_n$, $n\in \{1,2,3,4,6\}$
we arrive at the condition 
\eqn\ebound{
\nu(G)\  c_1(\cx L) \leq\;  \eta \ \ (\leq 12 c_1(\cx L)\ ),
} where $\nu(G)$ is the singularity dependent quantity
\eqn\nudef{
\nu(G)= max({(6d_n-n)\over d_n})\ ,
}
for all $n\in \{1,2,3,4,6\}$ with $d_n \neq 0$. Note that this argument also 
tells us that the minimal bundle $\cx N_{min}$ is always a power of
$\cx L$, a fact that is non-trivial for $h^{1,1}(B)>1$.
We have also indicated 
in \ebound\ an upper bound for $\eta$ of a very different origin.
It is not related to the 
stability of $\cx N$ but rather to the stability of $\eta^\prime=c_1(
\cx N^\prime)\geq 0$, where $\cx N^\prime$ is
the equivalent bundle in the second $E_8$ factor. 
 
{}From eqs. \ebound,\nudef\  we obtain 

\eqn\eboundsii{
\vbox{
\offinterlineskip
\tabskip=0pt\halign{\strut
\hfil~$#$&$\quad#$~\hfil\qquad\qquad&\hfil~$#$&
$\quad#$~\hfil\cr
 G=SU(2)^*:&  \eta\geq {14 \over 3} \cdot L& G=E_8:&  \eta\geq {0}\cdot L\cr
 G=SU(3)^*:&  \eta\geq {9 \over 2} \cdot L& G=E_7:&  \eta\geq {2}\cdot L\cr
 G=SU(M):&  \eta\geq {5} \cdot L& G=E_6:&  \eta\geq {3}\cdot L\cr
 G=SO(N):&  \eta\geq {4} \cdot L& G=G_2:&  {\eta\geq 4}\cdot L\cr
& & G=F_4:&  \eta\geq {3}\cdot L\cr
}}}
\ni
Here $N\in\{7,\dots,12\},\  M\in\{2,\dots,6\}$ and 
the groups with a star denote the singularities associated to $III$ and $IV^s$
fibers in the list of \Betal.
Moreover $L=c_1(\cx L)$. Note that our arguments for the bound \ebound\ give
a necessary but in general not sufficient criterion for the stability of the 
bundle. In particular the distinction between singularities associated to 
simply laced (SL) and non-simply laced (NSL) groups is often the 
specific form of the sections $f_{c,d}$ in the SL case rather than 
a difference in the topological class \Betal. Our arguments give
a criterion for the existence of a holomorphic section $f_{c,d}$ which 
is sufficient for SL groups, but it does not ensure that one gets a generic
enough section to describe also the NSL case. Thus the bounds given in 
\eboundsii\ are strict lower bounds only for the simply laced case.

It would be nice to have a derivation of the bound on $\eta$ that
is directly related to the structure group $H$ and in particular
also applies to the non-compact case with arbitrary rank.
A reasonable proposal based on \eboundsii\ is

\eqn\eboundsiii{
\vbox{
\offinterlineskip
\tabskip=0pt\halign{\strut
\hfil $#$&\quad $#$~\hfil\qquad\qquad&\hfil~$#$&
$\quad#$~\hfil\cr
 H=SU(N):&  \eta \geq {N} \cdot L& H=E_8:& \eta \geq  {5}\cdot L\cr
 H=SO(7):&  \eta \geq {4} \cdot L& H=E_7:& \eta \geq  {14 \over 3}\cdot L\cr
 H=SO(M):&  \eta \geq {M\over 2} \cdot L& H=E_6:& \eta \geq  {9 \over 2}\cdot L\cr
 H=Sp(K):&  \eta \geq {2K} \cdot L&H=G_2:& \eta \geq  {7 \over 2}\cdot L\cr
&&H=F_4:& \eta \geq  {13 \over 3}\cdot L\ ,\cr}}}
\ni
where $N\geq2$, $M\geq8$ and $K\geq 2$. Here we have used the calculation
of the six-dimensional spectrum in Table 3 of \Betal\ to raise the 
bounds for the NSL cases, 
such that we get the correct answer in six
dimensions. Similarly we could compare the calculation of the matter
spectrum in four dimension for a heterotic compactification on 
a Calabi--Yau 3-fold with \eboundsii. 

It is interesting to observe, how the geometric type IIA description 
contains non-trivial information about the instanton dynamics on 
Calabi--Yau manifolds $Z_n$. Moreover in many cases there is also another
point of view, namely in terms of a supersymmetric field theory 
in uncompactified space time. In particular, via the heterotic compactification on $K3\times T^2$,
the instanton dynamics on K3 is related to the moduli space of $\cx N=2$ supersymmetric 
field theories. Similarly, instantons on Calabi--Yau three-folds will be
related to the moduli space of $\cx N=1$ Super-Yang-Mills theories. It would be very
interesting to use the toric type IIA  construction of $V$ on $Z_n$ to study this relation 
in detail.

\vskip 1cm

\ni
{\bf Acknowledgements:}\br
P.M. would like to thank Govindan Rajesh for discussions. 
P.B. was supported in part by
the Natural Science Foundation under Grant No. PHY94-07194. P.B. would
also like to acknowledge  LBL,
Berkeley for hospitality during the course of this work.

\listrefs
\end